\documentclass[aps,prb,twocolumn,showpacs,superscriptaddress,groupedaddress]{revtex4}  % for review and submission
\usepackage{graphicx}% Include figure files
\usepackage{amsmath}% Include figure files
\usepackage{dcolumn}% Align table columns on decimal point
\usepackage{bm}% bold math

\begin{document}

\title{Paramagnetic Meissner, vortex and 'onion' ground states in Fulde-Ferrell finite-size superconductor}

\author{V. D. Plastovets}
\email{plastovec26@gmail.com}

\affiliation{Institute for Physics of Microstructures, Russian
Academy of Sciences, 603950, Nizhny Novgorod, GSP-105, Russia}

\affiliation{Lobachevsky State University of Nizhny Novgorod,
Nizhny Novgorod, 603950 Russia}

\affiliation{Sirius University of Science and Technology, 1
Olympic Ave, 354340 Sochi, Russia}

\author{D. Yu. Vodolazov}
\email{vodolazov@ipmras.ru}

\affiliation{Institute for Physics of Microstructures, Russian
Academy of Sciences, 603950, Nizhny Novgorod, GSP-105, Russia}

\date{\today}

\begin{abstract}

We theoretically find that finite size Fulde-Ferrell (FF)
superconductor (which is characterized by spatially nonuniform
ground state $\Psi \sim \text{exp}(-i{\bf q}_{FF}{\bf r})$ and
$|\Psi|(r)=const$ in the bulk case, where $\Psi$ is a
superconducting order parameter) has paramagnetic Meissner, vortex
and 'onion' ground states with $|\Psi|(r) \neq const$. These
states are realized due to boundary effect when the lateral size
of superconductor $L \sim 1/q_{FF}$. We argue, that predicted
states could be observed in thin disk/square made of
superconductor-ferromagnet-normal metal trilayer with $L \simeq
150-600 nm$.

\end{abstract}

\maketitle

\section{Introduction}

Majority of superconductors expel weak enough external magnetic
field which means that they are diamagnets. Mathematically this
property of superconductor can be described via London relation
between vector potential ${\bf A}$ and superconducting current
density ${\bf j}_s$:
\begin{equation}\label{eq1}
 {\bf j}_s=-\frac{c}{4\pi \lambda^2} {\bf A},
\end{equation}

where $\lambda$ is the London penetration depth which determines
how deep magnetic field penetrates the superconductor. However in
special case of so-called odd-frequency superconductivity
\cite{Bergeret_2001,Asano_2011,Yokoyama_2011,Walter_1998,Asano_2014,Alidoust_2014,Fominov_2015}
the sign in Eq. (1) may be opposite which corresponds to
paramagnetic response and formally negative density of Cooper
pairs $n\sim \lambda^{-2} <0$. Such a paramagnetic response can be
realized locally in different superconducting systems with
spatially nonuniform superconducting order parameter:
ferromagnetic (F) layer coupled to s-wave superconductor (S)
\cite{Bergeret_2001,Alidoust_2014,Fominov_2015} (as a practical
realization it could be ferromagnet/normal metal bilayer coupled
with s-wave superconductor \cite{Bernardo_2015}); normal metal (N)
layer coupled to p-wave superconductor \cite{Asano_2011} or to
s-wave superconductor with spin-active SN interface
\cite{Yokoyama_2011}; nonequilibrium N layer coupled to s-wave
superconductor \cite{Wilhelm_1995,Bobkova_2014}; near the edge of
clean p- or d -wave superconductor \cite{Walter_1998,Suzuki_2014}.

If thickness of the superconductor in hybrid SF or SN bilayers is
small the paramagnetic response of proximity induced odd-frequency
superconductivity in F or N layer may exceed the diamagnetic
response of the host superconductor which leads to global
paramagnetism. It may signal about instability and appearance of
modulated, along hybrid structure,
Fulde-Ferrell-Larkin-Ovchinnikov (FFLO) like superconducting state
\cite{Mironov_2012, Bobkova_PRB} which has zero magnetic response
($\lambda^{-2} \to 0$ in Eq. [\ref{eq1}]). In Larkin-Ovchinnikov
state the modulated state corresponds to the standing wave
(superconducting order parameter $\Psi \sim {\text cos}({\bf
q}_{LO}{\bf r})$) while in Fulde-Ferrell state $\Psi \sim {\text
exp}(i{\bf q}_{FF}{\bf r})$. Note that in the FF state spontaneous
currents can flow in the ground state \cite{Bobkova_PRB,
Mironov_2018} which are absent in LO state.
 \begin{figure}[]%1
 \includegraphics[width=1\linewidth]{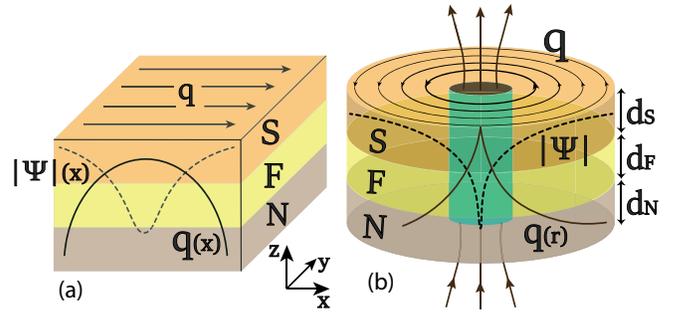}
 \caption{(a) Superconductor-ferromagnet-normal metal
 superconducting square being in quasi 1D Fulde-Ferrell state and
 (b) SFN disk being in vortex state. Dashed and solid curves
 correspond to distribution of $|\Psi|$ (the superconducting order
 parameter, averaged over the thickness) and $q$ ($\sim$
 supervelocity), respectively.} \label{fig:1}
 \end{figure}

In our work we theoretically study how ground state of FF
superconductor is modified when its lateral size becomes about of
$1/q_{FF}$. As a practical realization we have in mind square/disk
made of trilayer superconductor/ferromagnet/normal metal (see Fig.
\ref{fig:1}) where FF state can exist \cite{Mironov_2018}. We do
not consider LO state because in Ref. \cite{Marychev_2018} it has been 
found that in SFN trilayer FF state has lower energy than LO one.
Due to vanishing of normal component of superconducting current
$j_s$ at the border with vacuum in each (S, F and N) layer one has
to have $q = (\nabla \phi+2\pi A/\Phi_0)|_n=0$ ($\phi$ is the
phase of superconducting order parameter) which should affect the
classical Fulde-Ferrell (plain wave) state. In Refs.
\cite{Samokhin_2019, Plastovets_2019} it was found, that such a
modification occurs on scale about of $1/q_{FF}$ and it leads to
increase of the free energy $\textit{F}$. Therefore when $Lq_{FF}
\sim 1$ this increase of free energy may exceed the energy gain
from transition to modulated state and homogenous state becomes
more favorable. Support to this comes from Ref.
\cite{Samokhvalov_2010} where it was found that FFLO disk with
radius $Rq_{FF} \lesssim 1.2$ has higher critical temperature in
homogenous state than in FFLO one.

Our main result can be formulated as follows. We find that when
$Lq_{FF} \lesssim 2$ the ground state of FF superconductor is
spatially homogenous with $q=0$ and it has global paramagnetic
response ($\lambda^{-2}<0$) at small enough magnetic fields. With
increasing $L$ there is a second-order transition to quasi 1D
state with $\lambda^{-2}=0$ (see Fig. 1(a) or Fig. 2(a)) which
becomes energetically less favorable than single vortex state with
spatially dependent $\lambda^{-2}$ and $|\Psi|$ (see Fig. 1(b) or
Fig. 2(b)) at larger $L$. With further increasing $L$ the 'onion'
like state (see Fig. 2(c)) has the lowest energy.

In our calculations we use two models. Numerical solution of
Usadel equation for 3D square shown in Fig. \ref{fig:1} is rather
complicated and time-consumable problem. Therefore we use 2D
modified phenomenological Ginzburg-Landau (GL) equation (see
appendix A) for {\it thickness averaged} superconducting order parameter
$\Psi$ which qualitatively well describes main properties of FF
superconductor \cite{Samokhin_2017,Samokhin_2019,Plastovets_2019}
and allows us to find arbitrary in-plane distribution of $\Psi$,
$\lambda^{-2}$ and sheet current density. Besides we use 2D
Usadel equation (with radial and z dependence) for SFN disk 
where we consider only circularly symmetric (vortex)
and spatially homogenous states. The last approach allows us to
confirm results found from GL model and determine appropriate
material parameters for the experimental verification of the
predicted results.

\begin{figure}[]
\includegraphics[width=1\linewidth]{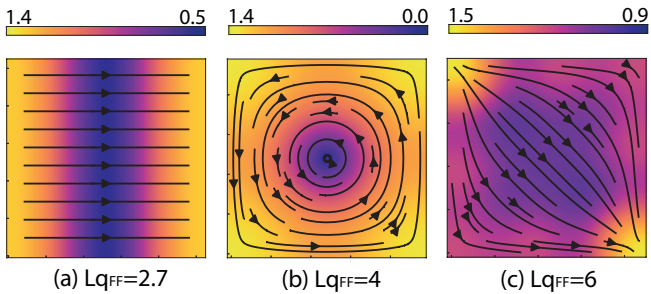}
\caption{Distribution of the dimensionless superconducting order
parameter $|\Psi|(x,y)$ and of ${\bf q}(x,y)$ (black arrows) in FF
square with different lateral size $L$. For all cases, the ground
states are shown (for $Lq_{FF} < \pi/\sqrt{2}$ ground state is the
homogenous state with $q=0$ - not shown here). (a) quasi 1D FF
like state with ${\bf q}=(q(x),0)$, (b) the single vortex state,
(c) the FF like state with ${\bf q}=(q_1(x,y), q_2(x,y))$. The
ground states are degenerate - the energy is not changed with
reversal of ${\bf q}$. } \label{fig:2}
\end{figure}

\section{Ginzburg-Landau approach}

First we present our results found from solution of modified 2D
Ginzburg-Landau equation (equation, boundary conditions,
parameters and numerical method are presented in appendix A). In Fig.
2 we show distribution of the superconducting order parameter
$|\Psi|$ and $q$ in FF superconducting square being in different
ground states. For $Lq_{FF} \leq \pi/\sqrt{2}$ the ground state
corresponds to spatially homogenous state ($q=0$) with
$\lambda^{-2}<0$. In narrow range $\pi/\sqrt{2}\leq Lq_{FF}
\lesssim 2.5$ the quasi 1D Fulde-Ferrell like state with
$\lambda^{-2}=0$, has the smallest energy (see Fig. 2(a)), while
for $2.5 \lesssim Lq_{FF} \lesssim 4.6$ the vortex state is a
ground one (see Fig. 2(b)) with $\lambda^{-2}(x,y)<0$ (see Fig. 3a
). For $Lq_{FF} \gtrsim 4.6$ the state with spatial distribution
of $|\Psi|$ and $q$ resembling the onion is realized (Fig. 2(c)).
In 'onion' state the diagonal distribution of $q$ minimizes the
positive contribution to the free energy which comes from
vanishing of $q|_n$ at the edge \cite{Plastovets_2019}. In
homogenous and quasi-1D FF states the current density is equal to
zero, while in the vortex and 'onion' states it is finite (see
Fig. 3).

 \begin{figure}[h]%3
\center{\includegraphics[width=1\linewidth]{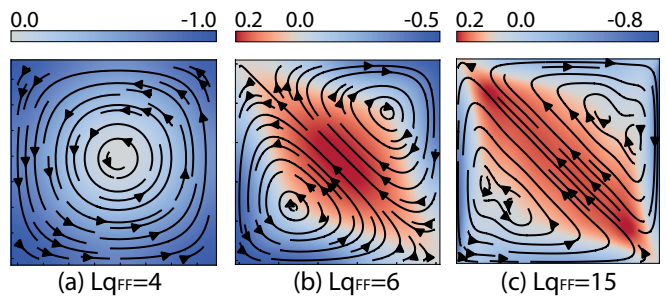}}
 \caption{Spatial distribution of local $\lambda^{-2}$ and
 superconducting current density (arrows) in squares being in
 vortex (a) and 'onion' (b,c) states. Red color corresponds to
 $\lambda^{-2}>0$ (diamagnetic response), blue one - $\lambda^{-2}
 < 0$ (paramagnetic response). One can see that in 'onion' state
 there are eddy currents, but vorticity $N=\oint\nabla \varphi
 \text{d{\bf l}}/2\pi$ is equal to zero. This result resembles the
 eddy currents in d- and p-wave mesoscopic disks \cite{Suzuki_2014}
 with the only qualitative difference that in 'onion' state eddy
 currents are finite even at zero magnetic field.}
 \end{figure}

In FF square various metastable states may exist and their number
increases with increasing of $L$ as in ordinary mesoscopic
superconductor (see for example \cite{Baelus_2004}). In Fig. 4 we
show examples of such states. These are vortex free state (not the
'onion' one - compare Fig. 4(a) and Fig. 2(c)) and states with
different number of vortices, including the states with vortices
and antivortices (Figs. 4(e,f)). In ordinary finite-size
(sometimes it is also called mesoscopic) superconductor vortex
states are stable in presence of the magnetic field $H$ (including
vortex-antivortex molecule \cite{Chibotaru_2000,Misko_2013}),
while in FF superconductor they are stable even at $H=0$. Their
stability comes from decreasing of the free energy and increase of $|\Psi|$
in presence of finite $q$ (which is proportional to supervelocity). In contrast, 
in ordinary superconductor finite $q$ leads to increase of $\textit{F}$ 
and suppression of superconductivity. The metastable and ground states are
degenerative - their energy does not change with reversal of $q$
and/or change the position of vortices. For instance, in case of
state shown in Fig. 4(b) there are three more states with the same
energy that can be realized by placing the vortex in different
quarters of the square.

\begin{figure}[h]
\includegraphics[width=1\linewidth]{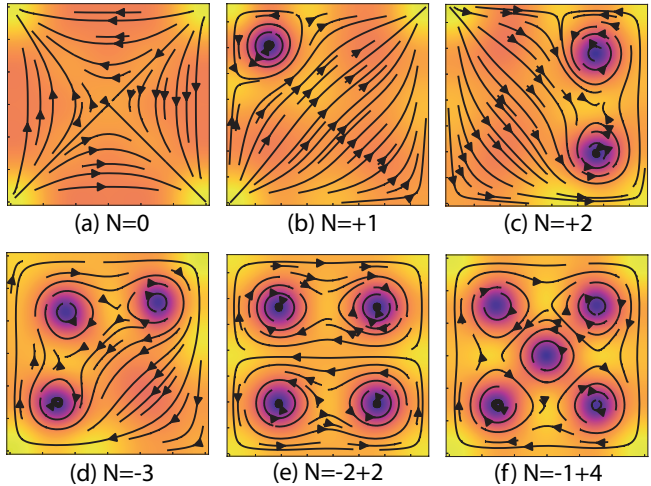}
\caption{Distribution of magnitude of the superconducting order
parameter $|\Psi|(x,y)|$ (colormap) and ${\bf q}(x,y)$ (black
arrows) in square with $Lq_{FF}=8$, being in various metastable
states. $N$ is a total vorticity ($N=\oint\nabla \varphi
\text{d{\bf l}}/\sqrt{2\pi}$, the contour ${\bf l}$ is positively
oriented along the boundaries).} \label{fig:4}
\end{figure}

In Fig. 5 we show dependence of the free energy $F$ and magnetic moment $M=-dF/dH$ of
FF square with the different $L$ on the magnetic field. Homogenous and 1D states 
demonstrate paramagnetic response at low fields (see Fig. 5(a,b)) which 
transforms to diamagnetic one at large fields. 
Because free energy decreases at low fields one can expect magnetic field
induced enhancement of critical temperature. Similar field induced
enhancement of $T_c$ was predicted for bulk FFLO superconducting
film being in parallel magnetic field
\cite{Bobkova_2014,Marychev_2018}, FFLO disk of small radius
\cite{Samokhvalov_2010} and 1D superconducting FFLO ring
\cite{Zyuzin_PRB} being in perpendicular magnetic field. Due to
small size of the square it cannot accommodate vortex even near
critical magnetic field which is also typical for ordinary small-size
superconductor. Because of that dependence $M(H)$ is not hysteretic.

Vortex (Fig. 5(c)) and 'onion' (Fig. 5(d)) states have more complicated $M(H)$ 
dependence because of entry/exit of vortices and antivortices. 
In Fig. 5(c) we show evolution of magnetic moment with increasing and 
decreasing of H for square being initially in vortex state. Perpendicular 
small magnetic field decreases q 
in the square and it results to increase of free energy and diamagnetic response. 
At $H\sim 0.4 H_{GL}$ vortex exits and antivortex enters, which considerably 
lowers the free energy and leads to the change of sign of $M$. It is interesting, 
that at $H \sim 1.5 H_{GL}$ antivortex exits and square goes to Meissner (vortex 
free) state which has the lowest energy in some range of magnetic fields 
among other states. 
It happens due to absence of the vortex core which increases the free energy 
while supervelocity is large enough due to the magnetic field. At $H \sim 2.5 H_{GL}$ 
the Meissner state becomes unstable and vortex enters the square. 
At $H \sim 3.2 H_{GL}$ square goes to
normal state. With decreasing of magnetic field the square passes through vortex, 
Meissner and finally, antivortex states. The dependence M(H) is 
hysteretic, as in case of ordinary mesosocopic superconductor, due to presence 
of energy barriers for vortex/antivortex entry/exit. Note, that in such a 
square one also can expect field enhanced critical temperature because 
free energy is lower at finite magnetic field than at $H=0$. 

Evolution of magnetic properties of the square being in 'onion' ground 
state at $H=0$ is more complex (see Fig. 5(d)). 
At $H\sim 0.15 H_{GL}$ there is first order transition from 'onion' 
to another vortex free state where magnetic response is paramagnetic 
(at lower fields magnetic response is weak and diamagnetic). 
At $H \sim 0.25 H_{GL}$ there is a transition to antivortex state with N=-1 
and then antivortex is replaced by a vortex (N=1) at $H\sim H_{GL}$.
The number of vortices is growing with increasing H to N=8 and, finally, 
square goes into the normal state.

\begin{figure}[h]
\includegraphics[width=1\linewidth]{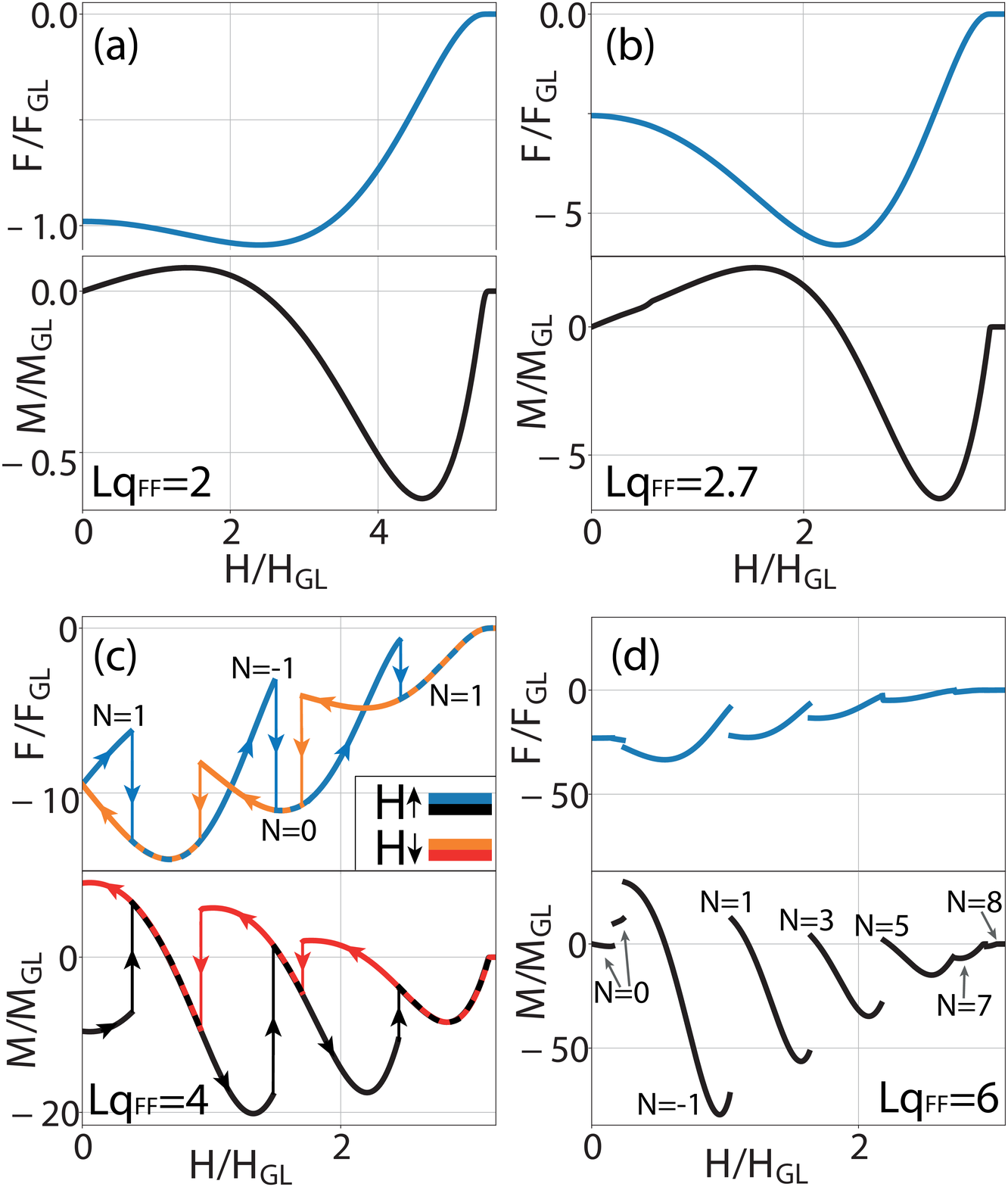}
\caption{Magnetic field dependence of the free energy 
and magnetic moment of FF squares with different lateral size shown in figures. 
In figures (a,b) $M(H)$ is reversable (no vortices at any $H$), 
while in figures (c,d) it is hysteretic 
(in (d) we show $M(H)$ only in increasing magnetic field). N is the vorticity 
(number of vortices). Free energy is 
scaled in units of $F_{GL}$ (see Appendix A), magnetic field is in units 
of $H_{GL}=\Phi_0/2\pi\xi_{GL}^2$ and magnetic moment is in units 
of $M_{GL}=F_{GL}/H_{GL}$.} \label{fig:5}
\end{figure}

\section{Usadel approach}

\begin{figure}[hbtp]%6
\includegraphics[width=0.5\textwidth]{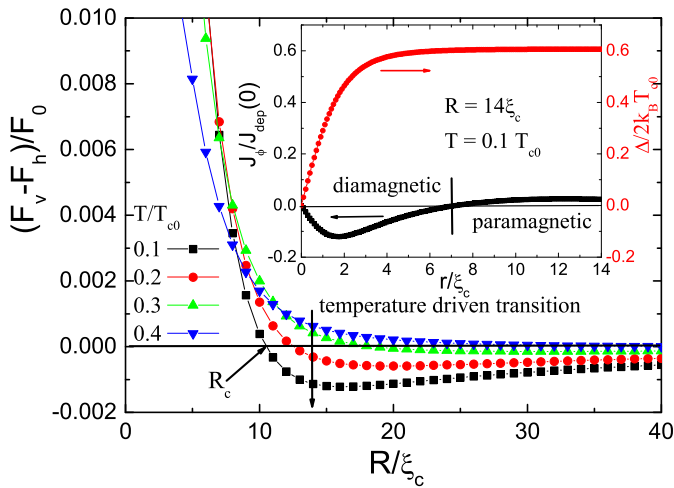}
\caption{Dependence of the difference between free energies of the
homogenous ($F_h$) and single vortex ($F_v$) states on radius of
SFN disk. In inset we show the radial dependence of sheet current
density and magnitude of superconducting order parameter in Usadel
model $\Delta$ in S layer on the boundary with vacuum. Parameters
of SFN disk: $d_S=1.4\xi_c$, $d_F=0.15 \xi_c$, $d_N=\xi_c$,
exchange energy in F layer is $E_{ex} =25 k_BT_{c0}$.}
\label{fig:6}
\end{figure}

Our calculations in framework of microscopic Usadel approach for
SFN disk supports some results found  in GL model (for details of
the method see Appendix B). We consider only circularly symmetric
states (homogenous and vortex ones) because finding of states
similar to ones shown in Fig. 2(a,c) and Fig. 4 needs solution of
3D Usadel equation, together with finding $q(\vec r)$. In Fig.
\ref{fig:6} we show difference between free energies of
homogenous and vortex states as a function of radius of the SFN
disk. Free energy (per unit of square) is normalized in units of
$F_0=\pi N(0)(k_BT_{c0})^2\xi_c$, where $\xi_c=(\hbar D_S/k_B
T_{c0})^{1/2}$, $N(0)$ is a one spin density of states in superconductor,
$T_{c0}$ is a critical temperature of single S layer and $D_S$ 
is its diffusion coefficient. One can see that starting from some radius $R_c$
(its value depends on temperature) vortex state becomes
energetically more favorable than homogenous one. Taking into
account that in bulk FF state $q_{FF}\simeq 0.16/\xi_c$  at $T=0.1
T_{c0}$ (see Fig. B.1) we estimate $2R_c \simeq 20 \xi_c \simeq
3.2/q_{FF}$ (for $T=0.3 T_{c0}$ we have $2R_c \simeq 38 \xi_c
\simeq 4.5/q_{FF}$) which is close to the numerical value,
obtained with help of modified GL equation for the square and to the
analytical result $2R_c \simeq 2.4/q_{FF}$ found from linearized
GL equation for the disk near $T_c$ (see Fig. 2 in
\cite{Samokhvalov_2010}). From Fig. \ref{fig:6} it also follows
that there exists temperature driven first order transition from
homogenous to vortex state as one decreases temperature.

In the vortex ground state there is a finite sheet current $J=\int
j_s{\text dz}$ - see inset in Fig. \ref{fig:6}. In contrast to the
vortex in ordinary superconductor the sheet current density
changes sign across the disk because in different parts of the
trilayer $\int\lambda^{-2}dz$ has different sign. Note that in framework of
GL model one has $\int \lambda^{-2}dz<0$ in the vortex
state everywhere in the FF square and there is no sign change of
J (see Fig. 3(a)). In SFN structure the size of the vortex core is
different in S (where it is about of $\xi_c$) and N (where it is
about of $\xi_N =(\hbar D_N/k_B T)^{1/2} \gg \xi_c$) layers. This
difference has been observed recently in SN bilayer
\cite{Stolyarov_2018,Panghotra_2019}. Due to vanishing 
of superconducting order parameter
in the center of vortex core the proximity induced odd-frequency
superconductivity is suppressed in N layer on scale about of
$\xi_N$ around the vortex core and it leads to $\lambda^{-2}>0$
there. Apparently this effect is not caught by used
phenomenological GL model.

Vortex state is a double degenerative state because states with
opposite vorticity have the same energy (if one neglects
interaction of vortex induced magnetic field with ferromagnet
layer). Using typical parameters of NbN as a S layer (resistivity
$\rho_n=200 \mu \Omega \cdot \text{cm}$, diffusion coefficient
$D=0.5 \text{cm}^2/\text{s}$, $T_{c0}=10 \text{K}$, $\xi_c=6.4$
\text{nm}) and other parameters as in Fig. \ref{fig:6} we find
magnetic field in the center of vortex $\sim 2 \text{Oe}$ ($R=14
\xi_c$). This magnetic field is much smaller than thermodynamic
field of NbN: $H_c=\sqrt{4\pi N(0)(1.76k_BT_{c0})^2} \sim 10^3
\text{Oe}$ and it gives small contribution to the free energy
(smaller than $10^{-4}F_0$). Besides this field is too small to
affect magnetic properties of F layer but it is large enough to be
measured by SQUID magnetometer (especially if there is an array of
SFN disks).

We also find that energy of the giant vortex state with vorticity
$N \geq 2$ is larger than single vortex state when $R< 40 \xi_c$
due to larger vortex core.
We do not consider larger disks because we expect that 'onion'
state is more energetically favorable in such samples.

Making a hole in the center of SFN disk favors appearance of
vortex state because there is no positive contribution from the
vortex core to the free energy. For example for SFN ring with
parameters as in Fig. \ref{fig:6} and width $\xi_c$ the homogenous
state becomes unfavorable at $R \gtrsim 5 \xi_c$. For smaller
rings supervelocity $v_s \sim q \sim 1/R$ is too large and
homogenous state has smaller energy.

\begin{figure}[hbtp]%7
\includegraphics[width=0.48\textwidth]{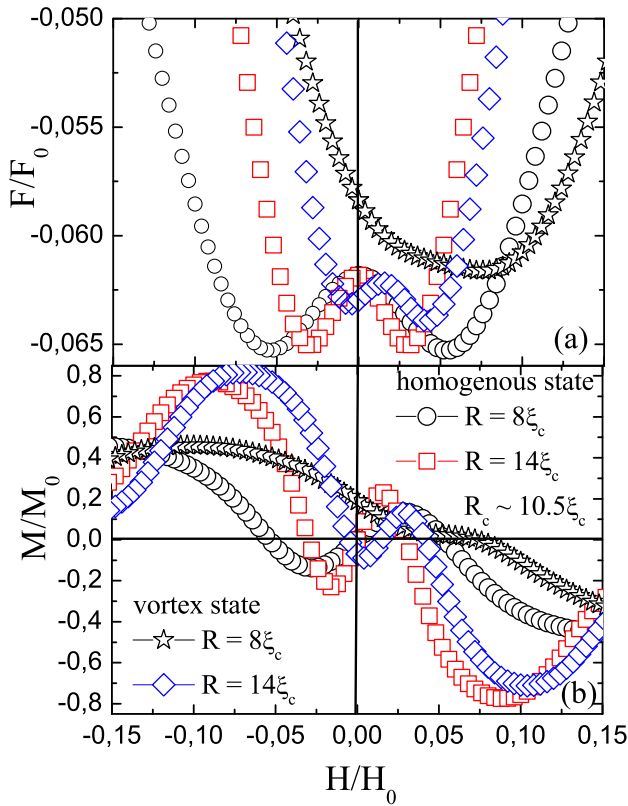}
\caption{(a) Dependence of the free energy of SFN disks being in
homogenous and vortex states on magnetic field
($H_0=\Phi_0/2\pi\xi_c^2 \sim 7.8 T$ for NbN). (b) Dependence of
the magnetic moment (measured in units of $M_0=2\pi
\xi_c^2R^2j_{dep}(0)/c$) of SFN disks on the magnetic field.
Parameters of the disks as in Fig. \ref{fig:6}.} \label{fig:7}
\end{figure}

In Fig. \ref{fig:7} we show calculated field dependencies of free
energy and magnetic moment of SFN disks with different radiuses
being either in vortex or Meissner states. In contrast to GL 
approach we cannot determine at which magnetic fields 
vortex/antivortex/Meissner states become 
unstable in the used model. Nevertheless these results 
support some conclusions following from GL model. Namely, disk being in
homogenous (Meissner) state has paramagnetic response up to some
magnetic field where it changes to ordinary diamagnetic
one (compare with Fig. 5(a,b)). It occurs roughly at  
$H\sim H^*\simeq \Phi_0/\xi_NR$ ($\xi_N=\sqrt{\hbar D_N/k_BT}$) 
when proximity induced superconductivity in N layer is got suppressed.

Evolution of states in SFN disk where vortex state is 
ground one at $H=0$ (see Fig. \ref{fig:7} for disk with $R = 14\xi_c$)
qualitatively resembles results found in GL model. Free energy increases 
at small fields and there is range of the fields where the Meissner state
has the lowest energy while at large fields vortex state becomes 
again more favorable. From Fig. \ref{fig:7} one may also conclude that 
antivortex state (it corresponds to vortex in negative magnetic 
field in Fig. \ref{fig:7}) decreases its energy at small $|H|$ 
which also coincides with our findings in GL model. 
Unusual behavior changes on conventional one at $H > H^*$ when proximity 
induced superconductivity dies in N layer.

Smaller size disk ($R=8\xi_c$) with vortex demonstrates different behavior
at low fields, but it is not clear if vortex state is stable there. 
We are only sure that at large fields ($H>H^*$), vortex state is more favorable
than Meissner one as in ordinary superconductors.

The similar results could be found for different material
parameters of SFN disk. For example one can take F layer with
larger or smaller exchange energy - it leads only to different
needed thickness of F layer when FF state could be realized (see
Fig. S4 in \cite{Mironov_2018}). One also may increase thickness
of N layer - it shifts the temperature where bulk FF state appears
(see Fig. S3(a) in \cite{Mironov_2018}). More crucial influence on
the predicted effect comes from increasing of $d_S$. While the
bulk FF state could be realized in case of relatively thick $d_S
\gtrsim 3 \xi_c$ (see Fig. S3(b) in \cite{Mironov_2018}) the
energy gain of bulk FF state in comparison with homogenous state
decreases with increasing $d_S$, as well as $|\lambda^{-2}|\to 0$
in homogenous state. As a result difference in energy between all
considered ground states decreases too and some of them could
disappear. For example in the vortex state large positive
contribution to $F$ comes from the vortex core and it could exceed
the negative contribution to $F$ due to finite $q$ around the
vortex.

\section{Discussion}

In contrast to other systems where {\it global} 'paramagnetic
Meissner effect' (PME) was theoretically or experimentally found
(high T$_c$ materials \cite{Walter_1998}, superconductors with
bulk pinning \cite{Thompson_1995,Koshelev_1995} or ordinary
mesoscopic superconductors \cite{Geim_1998}) in FF finite-size
superconductor it exists in the ground state when there is no any
frozen magnetic flux (trapped vortices). In this sense similar
global PME was also predicted for unconventional (p or d -wave)
superconducting disk where it may exist due to edge paramagnetic
currents \cite{Suzuki_2014}.

Ground and metastable states in SFN disk/square could be observed
with help of scanning tunnelling spectroscopy. In recent paper
\cite{Panghotra_2019} SN nanostructure (MoGe/Au) with needed
geometrical and physical parameters was studied. Indeed, Au is a
pure metal with residual resistance $\rho_N \simeq 2 \mu \Omega
\cdot \text{cm} $ even for 20 nm thick film \cite{Vries_1987},
while MoGe is rather dirty metal with $\rho_S \simeq 180 \mu
\Omega \cdot \text{cm} $, $T_{c0}=6.5 K$ and $D \simeq 0.4
\text{cm}^2/s$ \cite{Plourde_2001} ($\xi_c \simeq 6.8 \text{nm}$).
To observe predicted states one needs to add ferromagnet layer,
made from CuNi, for example, with thickness $d_F \sim 2-3
\text{nm}$ (it should be equal approximately to half of thickness
of ferromagnet in NbN/CuNi/NbN trilayer where $\pi$ Josephson
junction has been recently realized \cite{Yamashita_2017}). In homogenous state
superconducting order parameter is homogenous while in quasi 1D,
vortex and 'onion' states it depends on coordinate due to $q(x,y)$
(see Fig. 2). We also have to add, that spatial variation of local
density of states (LDOS) is rather different in S and N layers due
to different length scales ($\sim \xi_c$ in S layer and $\sim
\xi_N$ in N layer) leading, for example, to rather different size
of vortex core in S and N layers
\cite{Panghotra_2019,Stolyarov_2018} and this curcumstance should
be taken into account in the experiment.

From our calculations made in framework of Usadel model and
parameters of NbN or MoN it follows that lateral size of the
square/disk should be less than $20 \xi_c \simeq 130 nm$ (see Fig.
\ref{fig:6}) to observe paramagnetic Meissner state. Vortex state
exists in slightly larger squares/disks while 'onion' state could
be observed in samples with lateral size as large as 600 nm as it
follows from modified GL model (see Fig. 1(c) in \cite{suppl} and
for estimation we take $1/q_{FF} \sim 6 \xi_c$).

Depending on the parameters the FF phase may exist at
$T<T_c^{FFLO}=T_c$ or at $T<T_c^{FFLO}<T_c$ \cite{Mironov_2018}.
In first case $1/q_{FF}$ is finite at $T=T_c$ and, hence, at
$T<T_c$ the homogenous paramagnetic state may exist only in
superconductor of small size with $Lq_{FF} \lesssim 2-3$. In the
second case one can expect transition from the paramagnetic to
quasi 1D state shown in Fig. 2(a) (because it is second-order
transition), as one decreases temperature below $T_c^{FFLO}$
($1/q_{FF}$ decreases from infinity at $T=T_c^{FFLO}$ up to finite
value at lower temperature) but it may happen that for transition
to vortex or 'onion' state one needs to overcome the energy
barrier. If the energy barrier is too high the superconducting
square will stay in quasi 1D state. In this case one can use
perpendicular magnetic field to switch Fulde-Ferrell
superconductor from one state to another one, as it is 
demonstrated in section II.

We expect small influence of edge or bulk defects on our results due to 
large $\xi_N \gg \xi_c$. For considered above N layers $\xi_N > 50$ nm 
which is comparable with the size of SFN disk or square where these states 
could be observed.

\section{Summary}

We find that Fulde-Ferrell finite size superconductor may have
different ground states. Depending on the lateral size it could be
either in homogenous paramagnetic state, quasi 1D, vortex or
'onion' states. We propose that these states could be realized in
square(disk) made of superconductor/ferromagnet/normal metal
trilayer with lateral size 150-600 nm, where superconductor is
dirty superconducting material (like NbN, MoGe, NbTiN and so on),
normal metal is Au, Ag, Al or Cu and ferromagnet is CuNi or other
weak ferromagnetic material.

\begin{acknowledgements}

Authors acknowledge support from Foundation for the Advancement of
Theoretical Physics and Mathematics "Basis" (grant 18-1-2-64-2)
and Russian Foundation for Basic Research (project number
19-31-51019).

\end{acknowledgements}

\appendix

\renewcommand\thefigure{\thesection.\arabic{figure}}
\setcounter{figure}{0} %correct numeration of the figures in Appendix

\section{Modified Ginzburg-Landau model}

The modified Ginzburg-Landau free energy functional describing 2D
superconductor being in the Fulde-Ferrell-Larkin-Ovchinnikov
(FFLO) phase can be written as follows \cite{Buzdin_2007}

\begin{align}\label{eq:1}
F'=\alpha(T)|\Psi'|^2
+\frac{\beta}{2}|\Psi'|^4+\gamma(|\Pi_x\Psi'|^2+|\Pi_y\Psi'|^2)
\\ \notag
+\delta(|\Pi^2_x\Psi'|^2+|\Pi^2_y\Psi'|^2+|\Pi_x\Pi_y\Psi'|^2
+|\Pi_y\Pi_x\Psi'|^2)
\\ \notag
+\mu|\Psi'|^6,
\end{align}
where $\Psi'$ is a complex superconducting order parameter and $\Pi_{x,y}=\nabla_{x,y}-i(2e/\hbar c)A_{x,y}$. As in Refs.
\cite{Samokhin_2017,Samokhin_2019} we neglect term with
$|\Psi'|^6$ (it allows us to decrease the number of free
parameters) and define the signs of constants: $\alpha, \gamma<0$
and $ \beta, \delta >0$ to have Fulde-Ferrell state as a ground
one.

Ginzburg-Landau functional in the form similar to Eq. (A1) was
derived from microscopic theory for clean thin superconducting
film placed in parallel magnetic field \cite{Buzdin_1997}. We use
here similar GL functional to model properties of
superconductor/ferromagnet/normal metal trilayer being in
Fulde-Ferrell state, where superconductor and ferromagnet are
dirty metals with large resistivity. Therefore coefficients
$\alpha, \beta, \delta$, $\gamma$ should be considered only as a
phenomenological parameters and $\Psi'$ is a superconducting order
parameter averaged over the thickness of SFN trilayer.

The dimensionless free energy $F$ and order parameter $\Psi$ are
introduced as: $F'=F_{GL}F=(\alpha^2/\beta) F$, $\Psi'=\Psi_0
\Psi=\sqrt{|\alpha|/\beta} \Psi$, with defining of the
characteristic length $\xi_{GL}=\sqrt{|\gamma|/|\alpha|}$ and the
dimensionless parameter $\zeta=|\alpha|\delta/|\beta|^2$. Varying
$\int F{\text dS}$ with respect to $\Psi^*$, we obtain the
modified Ginzburg-Landau equation for the dimensionless order
parameter:
\begin{gather}\label{eq:2}
\zeta \{\Pi^4_x+\Pi^2_x\Pi^2_y+\Pi^2_y\Pi^2_x+\Pi^4_y\}\Psi
\\ \notag
+\{\Pi^2_x+\Pi^2_y\}\Psi+\Psi|\Psi|^2-\Psi=0.
\end{gather}

Equation (\ref{eq:2}) has to be supplemented by the boundary
conditions
\begin{gather}\label{eq:5}
\Pi \Psi\Big|_n=0, \quad \Pi^3 \Psi\Big|_n=0.
\end{gather}

Our choice of boundary conditions provides vanishing of normal
component of superconducting current $j_s|_n$ and $q|_n=(\nabla
\phi-(2e/\hbar c)A)|_n$ on the boundary of superconducting square with vacuum.
In ordinary Ginzburg-Landau model they vanish simultaneously if
one chooses $\Pi \Psi|_n=0$ while in modified GL model one
needs two conditions due to higher order of derivatives in Eq.
(1). Vanishing of $q|_n$ follows from the microscopic Usadel model
for SFN trilayer, when superconducting current in each layer has
to be equal to zero on boundary with vacuum. In framework of
modified GL model one deals with averaged over the thickness
current density $j_s$, $\lambda^{-2}$ and $\Psi$. In contrast to
ordinary GL model it could be the situation when $\lambda^{-2}=0$
and, hence, $j_s|_n \sim -\lambda^{-2} q|_n=0$ while $q|_n \neq 0$
which contradicts microscopic results for SFN trilayer. It is the
reason why we use boundary condition $\Pi\Psi |_n =0$. After
making this choice vanishing of $j_s|_n$ automatically leads to
$\Pi^3 \Psi|_n=0$.

In numerical calculations we use relaxation method with adding of
the time derivative $\partial \Psi/\partial t$ in the right hand
side of Eq. (\ref{eq:2}) and looking for $\Psi(x,y)$ which does
not depend on time. In calculations we use two values of parameter
$\zeta=0.5$ and 2. In both cases we obtain nearly the same
results.

The transition from the homogenous to quasi 1D state is the second
order phase transition as one increases L. In framework of used
model one can find analytically the critical size $L_c$ when it
occurs. By seeking the solution of Eq. (A2) in the following form:
$\Psi=(1+|\delta \Psi|)exp(i\phi)$ and assuming that $|\delta
\Psi|\ll 1$ and $\nabla \phi \ll 1$ one can find that
quasi 1D state may exist when $L>L_c=\pi
\sqrt{\zeta}=\pi/\sqrt{2}q_{FF}$ where $q_{FF}=1/\sqrt{2\zeta}$ is
the optimal $q$ which minimizes the free energy of FF bulk
superconductor.

\section{Uzadel model}

To calculate superconducting properties of SFN disk we use 2D
Usadel equation for anomalous $F=\sin \Theta$ and normal $G=\cos
\Theta$ Green functions in polar coordinates

\begin{equation}%
\begin{split}
& \hbar D \left(\frac{\partial^2\Theta}{\partial
z^2}+\frac{1}{r}\frac{d}{dr}r\frac{d\Theta}{dr}\right)- \\
& -\left(2(\hbar\omega_n+iE_{ex})+\hbar Dq^2\cos
\Theta\right)\sin \Theta+2\Delta \cos \Theta=0,
\end{split}
\end{equation}

where $D$ is a diffusion coefficient ($D=D_S, D_F, D_N$ in
superconducting, ferromagnet and normal layers, respectively),
$E_{ex} \neq 0$ is the exchange energy which is nonzero only in F
layer, $\hbar \omega_n=\pi k_BT(2n+1)$ is the Matsubara frequency,
$q=\nabla \phi+(2\pi/\Phi_0)A(r)$ ($A$ is a tangential
component of vector potential ${\bf A}=(0,Hr/2,0)$, $\nabla \phi = N/r$,
$N$ is a vorticity), $\Delta$ is a magnitude of superconducting
order parameter which should be found in Usadel model with help of
self-consistency equation
\begin{equation}%
\Delta \ln\left(\frac{T}{T_{c0}}\right)+2\pi k_B T\sum_{\omega_n
\geq 0} \left( \frac{\Delta}{\hbar\omega_n}-\sin \Theta\right)=0.
\end{equation}

We assume that in F and N layers $\Delta=0$ because of zero BCS
coupling constant there. $T_{c0}$ in Eq. (B2) is the critical
temperature of single S layer. We consider disk with thickness
$d_S+d_F+d_N \ll \lambda$ and radius $R \ll
\lambda^2/(d_S+d_F+d_N)$. Therefore we neglect corrections to
$A(r)$ which comes from superconducting currents.

Averaged over the thickness of the disc local $\lambda^{-2}$ and
superconducting current density are calculated as
\begin{equation}%
\begin{split}
\lambda^{-2} =\frac{16 \pi^2k_BT}{\hbar c^2 (d_S+d_F+d_N)} &\times \\
\times\int_0^{d_S+d_F+d_N}&\frac{1}{\rho}\sum_{\omega_n \geq 0} Re(
\sin^2\Theta) dz \\
j_s =-q\frac{c\Phi_0}{8\pi^2} \lambda^{-2}
\end{split}
\end{equation}

with resistivity $\rho=\rho_S, \rho_F, \rho_N$ in S, F and N
layers, respectively.

\begin{figure}[h]%
\includegraphics[width=0.5\textwidth]{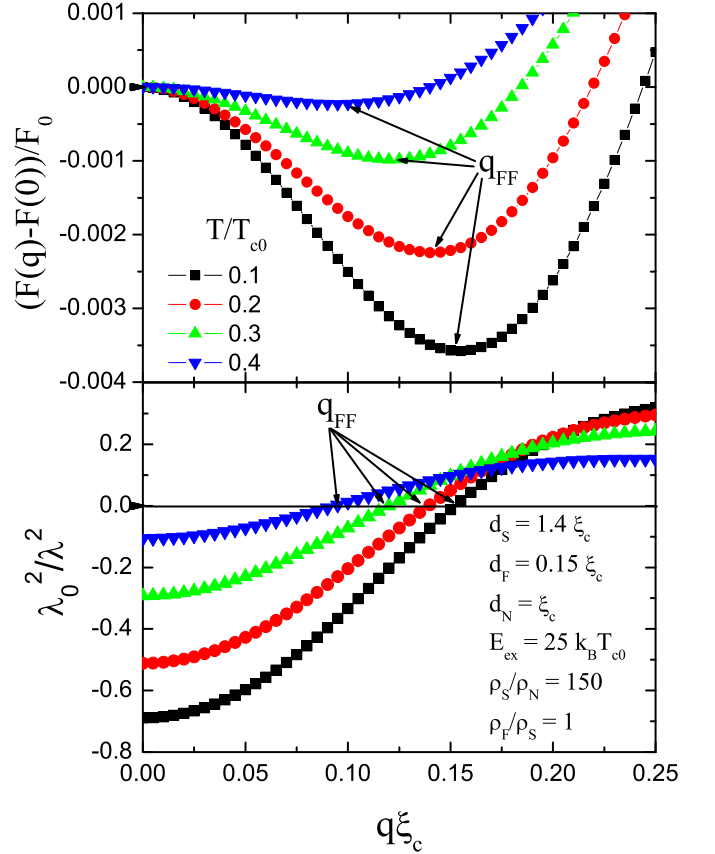}
\caption{Dependence of the free energy (a) and averaged over the
thickness of bulk SFN trilayer $\lambda^{-2}$ (b) on q.
$\lambda^{-2}=0$ in minimum of $F(q)$. Position of minimum shifts
to smaller $q$ with increasing temperature. For chosen parameters
$T_c=T_c^{FFLO}$.}  \label{fig:B.1}
\end{figure}

At boundaries between different layers we use Kupriyanov-Lukichev
boundary conditions \cite{KL} (for example
$D_Sd\Theta/dz=D_Fd\Theta/dz$ at $z=d_S$) with fully transparent
interfaces leading to continuity of $\Theta$. On the interface
with vacuum we use $d\Theta/dn=0$. Vanishing of normal component
of superconducting current density (not averaged over thickness)
on the boundaries with vacuum leads to $q|_n=0$. For considered
homogenous and single/giant vortex states it is fulfilled
automatically. In the center of the disk we use $\Theta(r=0)=0$ 
for vortex state and $d\Theta/dr|_{r=0}=0$ for homogenous one.

Equations (B1,B2) are solved numerically by using iteration
procedure. For initial distribution $\Delta(z,r)$ we solve Eq. (B1)
for Matsubara frequencies ranging from n=0 up to n=100. In
numerical procedure we use Newton method combined with tridiagonal
matrix algorithm. Found solution $\Theta(z,r)$ is inserted to Eq.
(B2) to find $\Delta(z,r)$ and than iterations repeat until the
relative change in $\Delta(z,r)$ between two iterations is 
larger than $10^{-8}$. Length is normalized in units of $\xi_c=(\hbar
D_S/k_BT_{c0})^{1/2}$, $\hbar \omega_n$, $\Delta$ is in units of
$k_BT_{c0}$, free energy (per unit of square) is normalized in
units of $F_0=\pi N(0)(k_BT_{c0})^2\xi_c$, current density is in
units of depairing current density of S-layer, magnetic field is
in units of $H_0=\Phi_0/2\pi\xi_c^2$ and $\lambda$ is in units of
magnetic field penetration depth in S layer $\lambda_0$. Step grid
in z direction is $dz=0.01-0.05 \xi_c$ (depending on the
layer) and in radial direction $dr =0.1 \xi_c$. We check that
results vary slightly with variation of $dz, dr$ if they are small
enough.

To decrease the number of free parameters we assume that the
densities of states in S, F and N layers are the same and ratio of
resistivities is equal to inverse ratio of diffusion constants or
mean path lengths $\rho_S/\rho_N=D_N/D_S=\ell_N/\ell_S$.

In Fig. B.1 we show dependence of free energy (it is calculated
using Eq. (S5) in \cite{Eltschka}) and $\lambda^{-2}$ on $q$ for
bulk SFN trilayer. For chosen parameters $T_c^{FFLO}=T_c \simeq
0.54 T_{c0}$. The minimum of $F$ is reached at $q=q_{FF}$ which
depends on temperature.

\end{document}